\documentclass[usenatbib]{pasj01}
\usepackage{natbib}

\begin{document} 
\Received{2018/06/30}
\Accepted{2019/01/21}

\title{Infrared luminosity functions based on 18 mid-infrared bands: revealing cosmic star formation history with AKARI and Hyper Suprime-Cam\footnote{Based in part on data collected at Subaru Telescope, which is operated by the National Astronomical Observatory of Japan.}}

\author{Tomotsugu \textsc{Goto}\altaffilmark{1}%
}
\altaffiltext{1}{National Tsing Hua University, No. 101, Section 2, Kuang-Fu Road, Hsinchu, Taiwan 30013}
\email{tomo@gapp.nthu.edu.tw}
\author{Nagisa  \textsc{Oi}\altaffilmark{2}}
\altaffiltext{2}{Tokyo University of Science, 1-3 Kagurazaka, Shinjuku-ku, Tokyo 162-8601, Japan}
\author{Yousuke  \textsc{Utsumi}\altaffilmark{3}}\altaffiltext{3}{
Kavli Institute for Particle Astrophysics and Cosmology (KIPAC), SLAC National Accelerator Laboratory, Stanford University, SLAC, 2575 Sand Hill Road, Menlo Park, CA  94025, USA}
\author{Rieko  \textsc{Momose}\altaffilmark{1,4}}\altaffiltext{4}{
Department of Astronomy, School of Science, The University of Tokyo
7-3-1 Hongo, Bunkyo-ku, Tokyo 113-0033, JAPAN}
\author{Hideo  \textsc{Matsuhara}\altaffilmark{5}}\altaffiltext{5}{Institute of Space and Astronautical Science, Japan Aerospace Exploration Agency, 3-1-1 Yoshinodai, Chuo, Sagamihara, Kanagawa 252-5210, Japan}
\author{Tetsuya  \textsc{Hashimoto}\altaffilmark{1}}
\author{Yoshiki  \textsc{Toba}\altaffilmark{6}}\altaffiltext{6}{Academia Sinica Institute of Astronomy and Astrophysics, P.O. Box 23-141, Taipei 10617, Taiwan}
\author{Youichi  \textsc{Ohyama}\altaffilmark{6}}
\author{Toshinobu  \textsc{Takagi}\altaffilmark{7}}\altaffiltext{7}{Japan Space Forum, 3-2-1, Kandasurugadai, Chiyoda-ku, Tokyo 101-0062 Japan}
\author{Chia Ying  \textsc{Chiang}\altaffilmark{1}}
\author{Seong Jin  \textsc{Kim}\altaffilmark{1}}
\author{Ece  \textsc{Kilerci Eser}\altaffilmark{1}}
\author{Matthew  \textsc{Malkan}\altaffilmark{8}}\altaffiltext{8}{Department of Physics and Astronomy, UCLA, Los Angeles, CA, 90095-1547, USA}
\author{Helen  \textsc{Kim}\altaffilmark{8}}
\author{Takamitsu  \textsc{Miyaji}\altaffilmark{9}}
\altaffiltext{9}{Insitituto de Astronom\'ia, Universidad Nacional Aut\'onoma de M\'exico
}
\author{Myungshin  \textsc{Im}\altaffilmark{10}}
\altaffiltext{10}{Astronomy Program, Department of Physics \& Astronomy, FPRD, Seoul National University, Shillim-Dong, Kwanak-Gu, Seoul 151-742, Korea}
\author{Takao  \textsc{Nakagawa}\altaffilmark{5}}
\author{Woong-Seob  \textsc{Jeong}\altaffilmark{11,12}}
\altaffiltext{11}{Korea Astronomy and Space Science Institute (KASI), 776 Daedeok-daero, Yuseong-gu, Daejeon 34055, Korea}
\altaffiltext{12}{Korea University of Science and Technology, 217 Gajeong-ro, 
Yuseong-gu, Daejeon 34113, Korea}
\author{Chris  \textsc{Pearson}\altaffilmark{13,14}}
\altaffiltext{13}{RAL Space, STFC Rutherford Appleton Laboratory, Didcot, Oxfordshire OX11 0QX, UK }
\altaffiltext{14}{The Open University, Milton Keynes, MK7 6AA, UK}
\author{Laia  \textsc{Barrufet}\altaffilmark{15}}
\altaffiltext{15}{European Space Astronomy Centre, 28691 Villanueva de la Canada, Spain}
\author{Chris  \textsc{Sedgwick}\altaffilmark{14}}
\author{Denis  \textsc{Burgarella}\altaffilmark{16}}
\author{Veronique \textsc{Buat}\altaffilmark{16}}\altaffiltext{16}{
Aix-Marseille Universit©¦©Õ, CNRS ¡ò©ñ LAM (Laboratoire d©£©¾Astrophysique de Marseille) UMR 7326, 13388 Marseille, France}
\author{Hiroyuki \textsc{Ikeda}\altaffilmark{17}}
\altaffiltext{17}{National Astronomical Observatory, 2-21-1 Osawa, Mitaka, Tokyo, Japan}

\maketitle

\begin{abstract}

 Much of the star formation is obscured by dust. For the complete understanding of the cosmic star formation history (CSFH), infrared (IR) census is indispensable.   AKARI carried out deep mid-infrared observations using its continuous 9-band filters in the North Ecliptic Pole (NEP) field (5.4 deg$^2$). This took significant amount of satellite's lifetime,  $\sim$10\% of the entire pointed observations.
 By combining archival Spitzer (5 bands) and WISE (4 bands) mid-IR photometry, we have, in total, 18 band mid-IR photometry, which is the most comprehensive photometric coverage in mid-IR for thousands of galaxies. However previously, we only had shallow optical imaging ($\sim$25.9ABmag) in a small area of 1.0 deg$^2$. As a result, there remained thousands of AKARI's infrared sources undetected in optical.
 Using the new Hyper Suprime-Cam on Subaru telescope, we obtained deep enough optical images of the entire AKARI NEP field in 5 broad bands ($g\sim$27.5mag). These provided photometric redshift, and thereby IR luminosity for the previously undetected faint AKARI IR sources. Combined with
 the accurate mid-IR luminosity measurement, we constructed mid-IR LFs, and thereby performed a census of dust-obscured CSFH in the entire AKARI NEP field. 
 We have measured restframe 8$\mu$m, 12$\mu$m luminosity functions (LFs), and estimated total infrared LFs at 0.35$<$z$<$2.2.  Our results are consistent with our previous work, but with much reduced statistical errors thanks to the large area coverage of the new data. We have possibly witnessed the turnover of CSFH at $z\sim$2.

\end{abstract}

\section{Introduction}


Mid-infrared (mid-IR) is one of the less explored wavelengths due to the earth's atmosphere, and difficulties in developing sensitive detectors. NASA's Spitzer and WISE space telescopes only had four filters in the mid-IR wavelength range, hampering studies of distant galaxies.

AKARI space telescope has a potential to revolutionize the field. Using its 9 continuous mid-IR filters (2-24$\mu$m), AKARI performed a deep imaging survey in the North Ecliptic Pole (NEP) field  over 5.4 deg$^2$. Using AKARI's 9 mid-IR band photometry, mid-IR SED diagnosis can be performed for thousands of galaxies, for the first time, over the large enough area to overcome cosmic variance.
Environmental effects on galaxy evolution can be also investigated with the large volume coverage \citep{2008MNRAS.391.1758K,cluster_LF}.

 \begin{figure*}
  \begin{center}
    \includegraphics[scale=0.4]{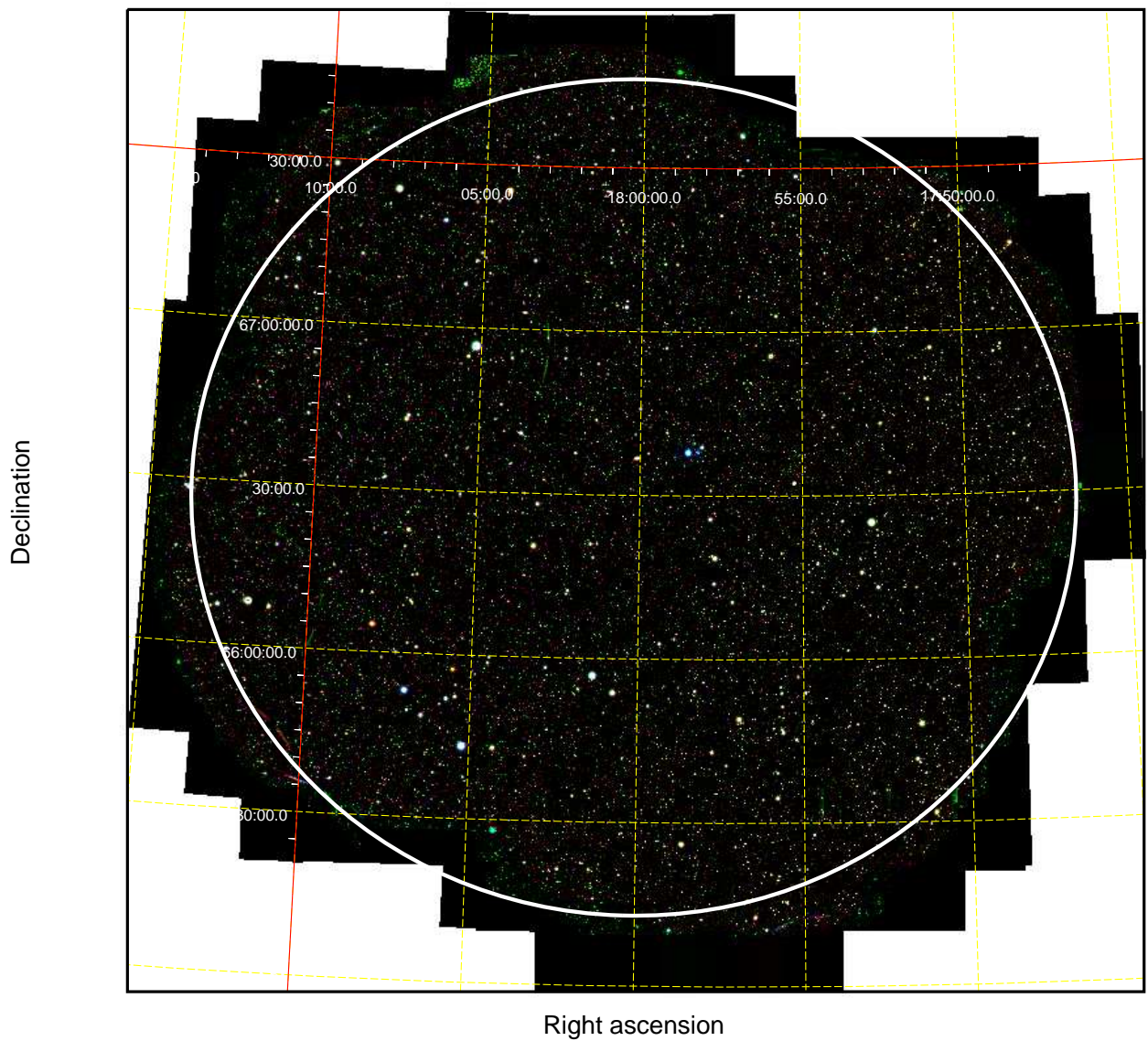}
  \caption{
       HSC three color ($g,r,i$) composite image of the NEP wide field (5.4 deg$^2$).
       The AKARI NEP wide data exist within the white circle. 
       }
  \label{fig:HSC}
  \end{center}
 \end{figure*}


 However, previously, we were limited by a poor optical coverage both in area and depths. Over this wide area,  only shallow optical/NIR imaging data have been available \citep{2007ApJS..172..583H,2010ApJS..190..166J,2014ApJS..214...20J}. Deep optical images are limited to the central 0.25 deg$^2$.

 To overcome these problems, we have newly obtained deeper optical data over the entire AKARI NEP wide field, using the Hyper-Suprime Cam on the Subaru telescope.
 Using the deeper optical data, in this paper, we measure  mid-infrared galaxy LFs,  and estimate total IR LFs (based on the mid-IR SED fitting) from the entire AKARI NEP field.
  Unless otherwise stated, we assume a cosmology with $(h,\Omega_m,\Omega_\Lambda) = (0.7,0.3,0.7)$.

\section{Data\label{sec:data}}

To rectify the situation and to fully exploit the AKARI's space-based data, we carried out an optical survey of the AKARI NEP wide field (PI:Goto) using Subaru's new Hyper Suprime-Cam \cite[HSC;][]{2018PASJ...70S...1M} in five optical bands ($g,r,i,z,$ and $y$, Oi et al. 2018 submitted).
  The HSC has a field-of-view (FoV) of 1.5 deg in diameter, covered with 104 red-sensitive CCDs.
  It has the largest FoV among optical cameras on 8m-class telescopes, and  can cover the AKARI NEP wide field (5.4 deg$^2$) with only 4 FoV (Fig.\ref{fig:HSC}).  The 5 sigma limiting magnitudes are 27.18, 26.71, 26.10, 25.26, and 24.78 mag [AB] in $g, r, i, z$, and $y$-bands, respectively.
See Oi et al. (2018, submitted) for more details of the observation and data reduction.

Subaru telescope does not have $u^*$-band capability, while it is critically important to accurately estimate photometric redshifts (photo-z) of low-z galaxies.
 Therefore, we obtained $u^*$-band image of the AKARI NEP wide field using the Megaprime camera of Canada France Hawaii Telescope \citep[PI:Goto,][]{GotoPKAS}. Combining the optical six bands, we have obtained accurate photo-z in the AKARI NEP field (Oi et al. 2018, submitted). To the detection limit in $L18W$ filter \citep[18.3 ABmag,][]{2012A&A...548A..29K}, we have 5078 infrared sources.

 In addition to the AKARI's 9 mid-IR bands, in the AKARI NEP field, there exit archival deep Spitzer \citep[IRAC1,2,3,4 and MIPS24,][]{2018ApJS..234...38N} and WISE ($W1,W2,W3,$ and $W4$) images as well. By combining all available mid-IR bands, in total we used 18 mid-IR bands, which are one of the most comprehensive mid-IR data sets for thousands of galaxies.


\section{Analysis\label{sec:ana}}


To compute LFs, we use the 1/$V_{\max}$ method, following \citet{GotoTakagi2010,GotoCFHT}.
 Uncertainties of the LF values include fluctuations in 
 the number of sources in each luminosity bin, 
 the photometric redshift uncertainties,
 the $k$-correction uncertainties,
 and the flux errors. 
 To estimate  errors, we used Monte Carlo simulations from 1000 simulated catalogs.
 Each simulated catalog contains the same number of sources.
 These sources are assigned with a new redshift, to follow a Gaussian distribution centered at the photo-z  with the width of $\Delta z/(1+z)$ ($\sim0.060$, Oi et al. in preparation).
 A new flux is also assigned following a Gaussian distribution with the width of flux error.
  For total infrared (TIR) LF errors, we re-performed the SED fit for the 1000 simulated catalogs.
    Note that total infrared luminosity is estimated based on mid-IR SED fitting although we have intensive 18-band filter coverage in mid-IR, as explained in Section 4.3.
 We ignored the cosmic variance due to our much improved volume coverage.
 All the other errors described above are added to the Poisson errors for each LF bin in quadrature.

\section{Results}
\label{results}

\subsection{The 8$\mu$m LF}\label{sec:8umlf}

\begin{figure}
 \includegraphics[height=8cm,trim={2cm 0 4cm 0},clip]{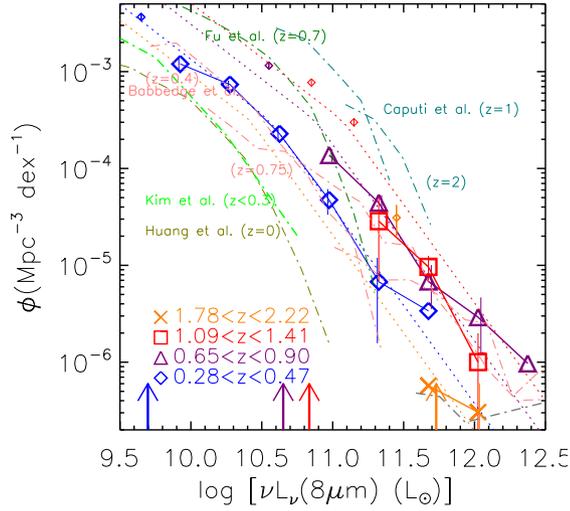}
 \caption{
  Restframe  8$\mu$m LFs based on the AKARI NEP wide field.
 The blue diamonds, the purple triangles, the red squares, and the orange crosses show the 8$\mu$m LFs at $0.28<z<0.47, 0.65<z<0.90, 1.09<z<1.41$, and $1.78<z<2.22$, respectively.
 AKARI's MIR filters can observe restframe 8$\mu$m at these redshifts in a corresponding filter. Error bars are estimated from the Monte Carlo simulations ($\S$\ref{sec:ana}).
 The dotted lines show analytic fits with a double-power law.
  The smaller data points at the faint ends are adopted from the NEP deep field, where AKARI data are deeper \citep{GotoCFHT}, and are included in the fit.
 Vertical arrows show the 8$\mu$m luminosity corresponding to the flux limit at the central redshift in each redshift bin.
 Overplotted LFs are  \citet{2006MNRAS.370.1159B} in the pink dash-dotted lines, \citet{2007ApJ...660...97C} in the cyan dash-dotted lines, \citet{2007ApJ...664..840H} in the dark-yellow dash-dotted lines, \citet{2010ApJ...722..653F},  in the dark green dash-dotted line, and \citet{Kimetal} in the bright green dash-dotted line. Best-fit parameters are presented in Table \ref{tab:fit_parameters}.
 }
   \label{fig:8um}
\end{figure}

 We first present monochromatic 8$\mu$m LFs, because the 8$\mu$m luminosity ($L_{8\mu m}$) has been known as a good indicator of the TIR luminosity \citep{2006MNRAS.370.1159B,2007ApJ...664..840H,Goto2011SDSS}.


An advantage of AKARI is that we do not need $k$-correction because one of the continuous filters always convert the rest-frame  8$\mu$m at our redshift range of $0.28<z<2.22$.
 Often in previous work, SED based  extrapolation was needed to estimate the 8$\mu$m luminosity. This was often the largest uncertainty.  This is not the case for the analysis present in this paper. 


 To estimate the restframe 8$\mu$m LFs, we followed our previous method in \citet{GotoCFHT} as we briefly summarize below.
  We used sources down to 80\% completeness limits \citep{2012A&A...548A..29K}.  
Galaxies are excluded when SEDs were better fit to QSO templates (2\%  from the sample). 

 We corrected for  the completeness using \citet{2012A&A...548A..29K} (25\% correction at maximum, with our selection to the 80\% completeness limits).
 Four redshift bins of 0.28$<z<$0.47, 0.65$<z<$0.90, 1.09$<z<$1.41,  and 1.78$<z<$2.22, were used, following our previous work. 
Then, the 1/$V_{\max}$ method was used to compensate for the flux limit.

The resulting restframe 8$\mu$m LFs are shown in Fig. \ref{fig:8um}. 
The arrows show the flux limit at the median redshift in bin.
We performed the Monte Carlo simulation to obtain errors. They are smaller than in our previous work \citep{GotoTakagi2010,GotoCFHT} thanks to the improved area coverage.
 The faint end marked with smaller data points are adopted from the NEP deep field, where AKARI data are deeper \citep{GotoCFHT}.

 Various previous studies are shown with dashed lines for comparison.
 Compared to the local LF, our  8$\mu$m LFs show strong evolution in luminosity up to $z\sim0.9$.
 Interestingly, the 8$\mu$m LFs peaks in the 3rd bin (z$\sim$1), then declines toward z$\sim$2.

\begin{figure}
 \includegraphics[height=8cm,trim={2cm 0 4cm 0},clip]{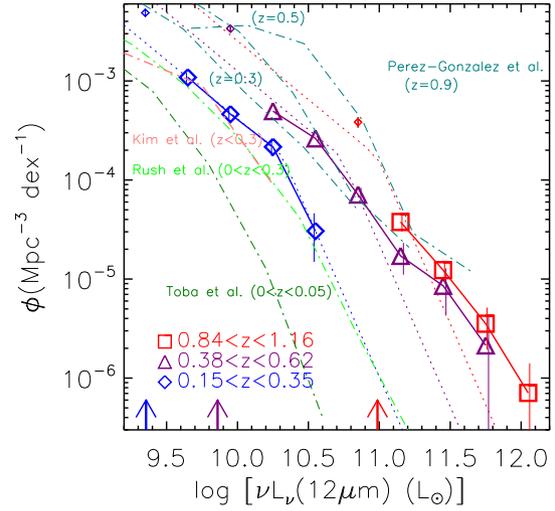}
 \caption{
 Restframe  12$\mu$m LFs based on the AKARI NEP wide field.
Luminosity unit is logarithmic solar luminosity ($L_{\odot}$).
  The blue diamonds, the purple triangles, and the red squares show the 12$\mu$m LFs at $0.15<z<0.35, 0.38<z<0.62$, and $0.84<z<1.16$, respectively.
The smaller data points at the faint ends are adopted from the NEP deep field, where AKARI data are deeper \citep{GotoCFHT}, and are included in the fit.
 Vertical arrows show the 12$\mu$m luminosity corresponding to the flux limit at the central redshift in each redshift bin.
  Overplotted LFs are  \citet{2005ApJ...630...82P} at $z$=0.3, 0.5 and 0.9 in the dark-cyan dash-dotted lines, 
\citet{2014ApJ...788...45T} at 0$<z<$0.05 based on WISE in the dark green dash-dotted lines,
  \citet{1993ApJS...89....1R} at 0$<z<$0.3 in the light green dash-dotted lines,
 and \citet{Kimetal}  at 0$<z<$0.3 in the pink dash-dotted line.
 Note \citet{1993ApJS...89....1R} is at higher redshifts than \citet{2014ApJ...788...45T}. Best-fit parameters are presented in Table \ref{tab:fit_parameters}.
 }
   \label{fig:12um}
\end{figure}

\subsection{12$\mu$m LF}\label{sec:12umlf}

Next, we show 12$\mu$m LFs. The 12$\mu$m luminosity $L_{12\mu m}$) is also known to correlate well with the TIR luminosity \citep{1995ApJ...453..616S,2005ApJ...630...82P}. 
AKARI's advantage still holds in not needing extrapolation based on SED models.
 The $L15,L18W$ and $L24$ filters cover the restframe 12$\mu$m at $z$=0.25, 0.5, and 1, respectively.
 Our analyses are the same with the 8$\mu$m LF; down to the 80\% completeness limit in each filter,  completeness correction, and the 1/$V_{\max}$  method.

Fig. \ref{fig:12um} shows the results. Various previous studies are shown in dash-dotted lines.
  Our 12$\mu$m LFs show steady evolution with increasing redshift.
 Similar to the 8$\mu$m  LF, the evolution becomes less evident between the two higher redshift bins.

 \subsection{Total IR LFs estimated from mid-IR SED fit} \label{sec:tirlf}

\begin{figure*}
\begin{center}
\begin{center}
 \includegraphics[height=8cm]{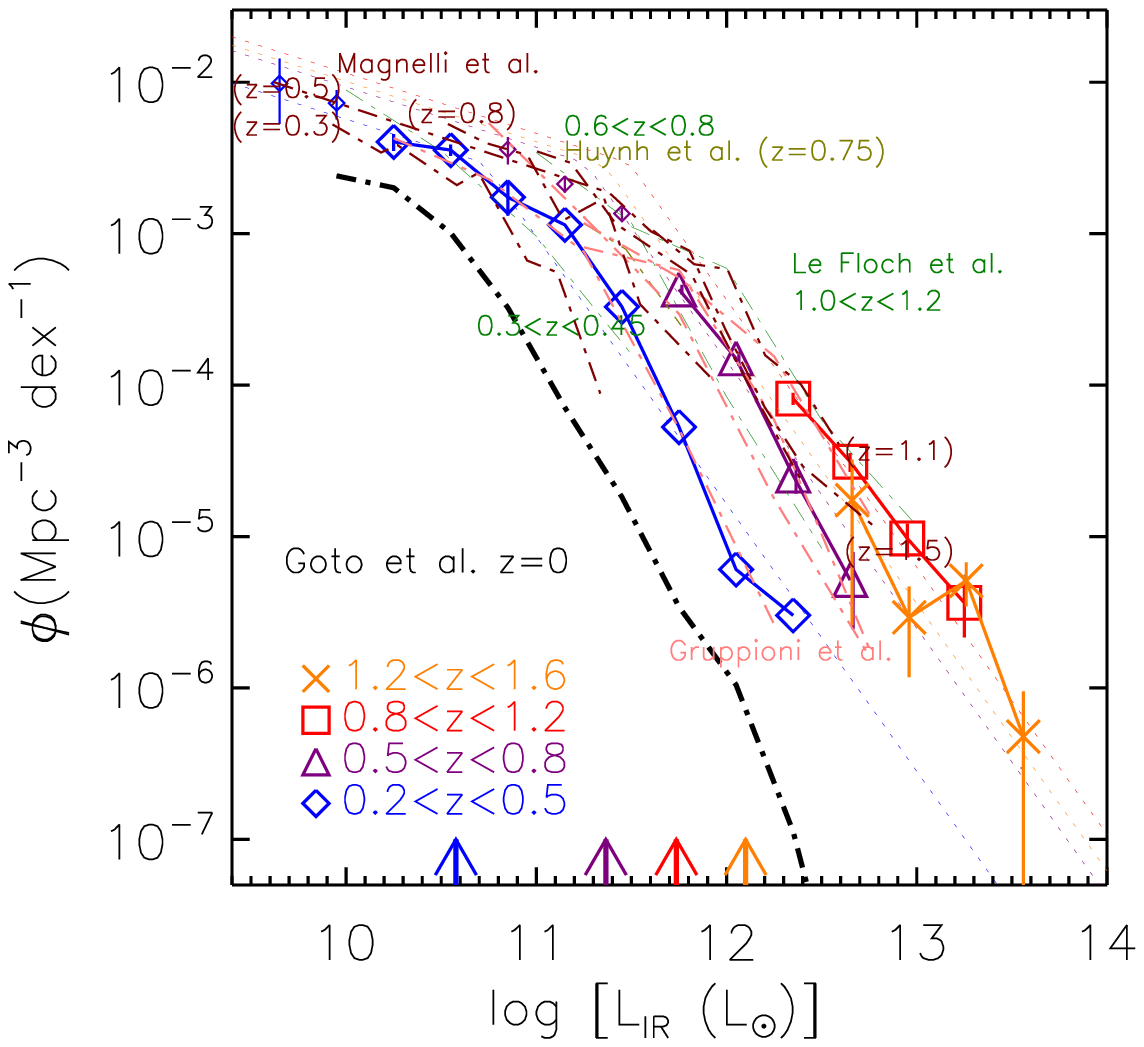}
 \end{center}
 \includegraphics[height=7cm,trim={2cm 0 4cm 0},clip]{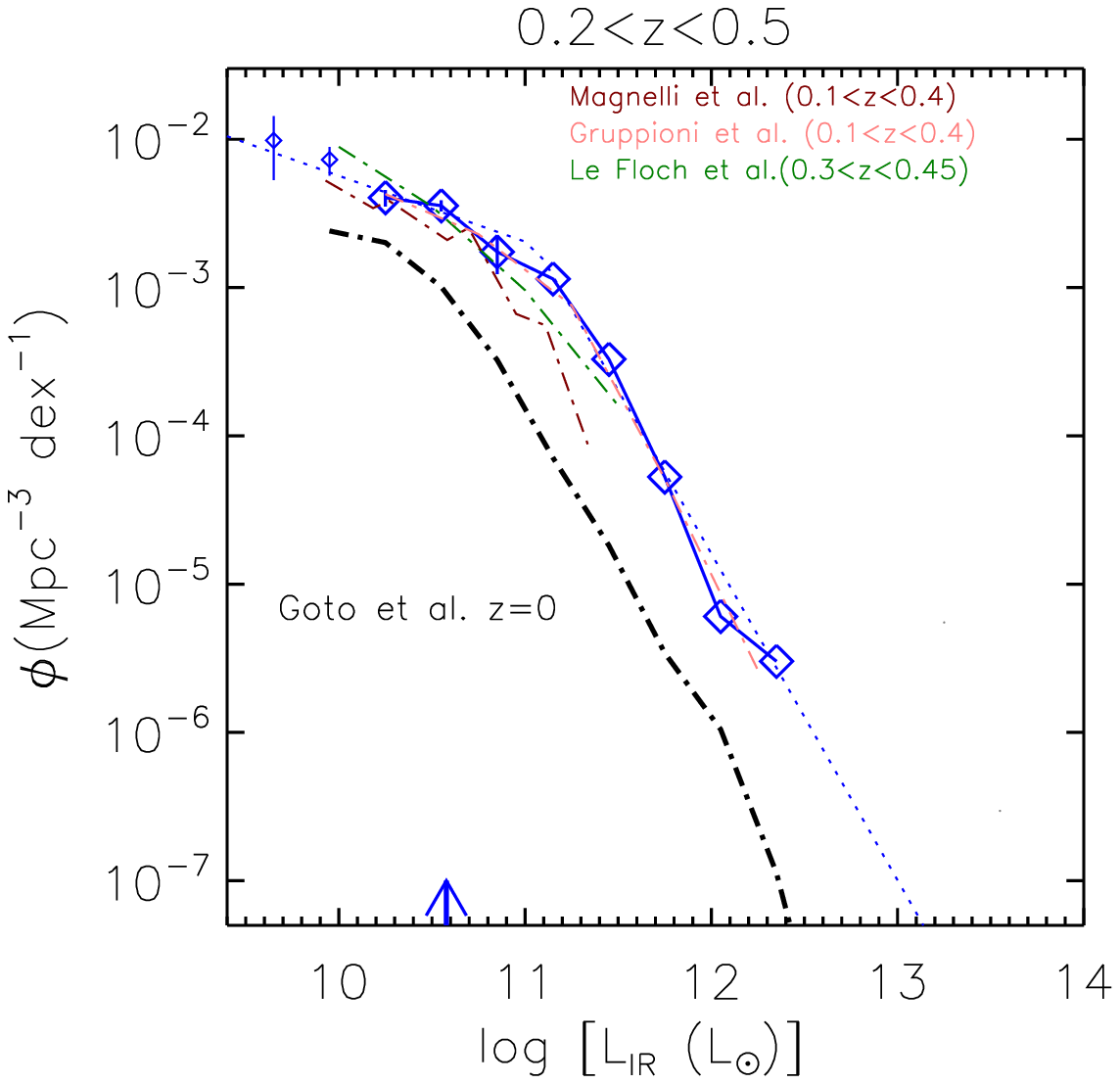}
 \includegraphics[height=7cm,trim={2cm 0 4cm 0},clip]{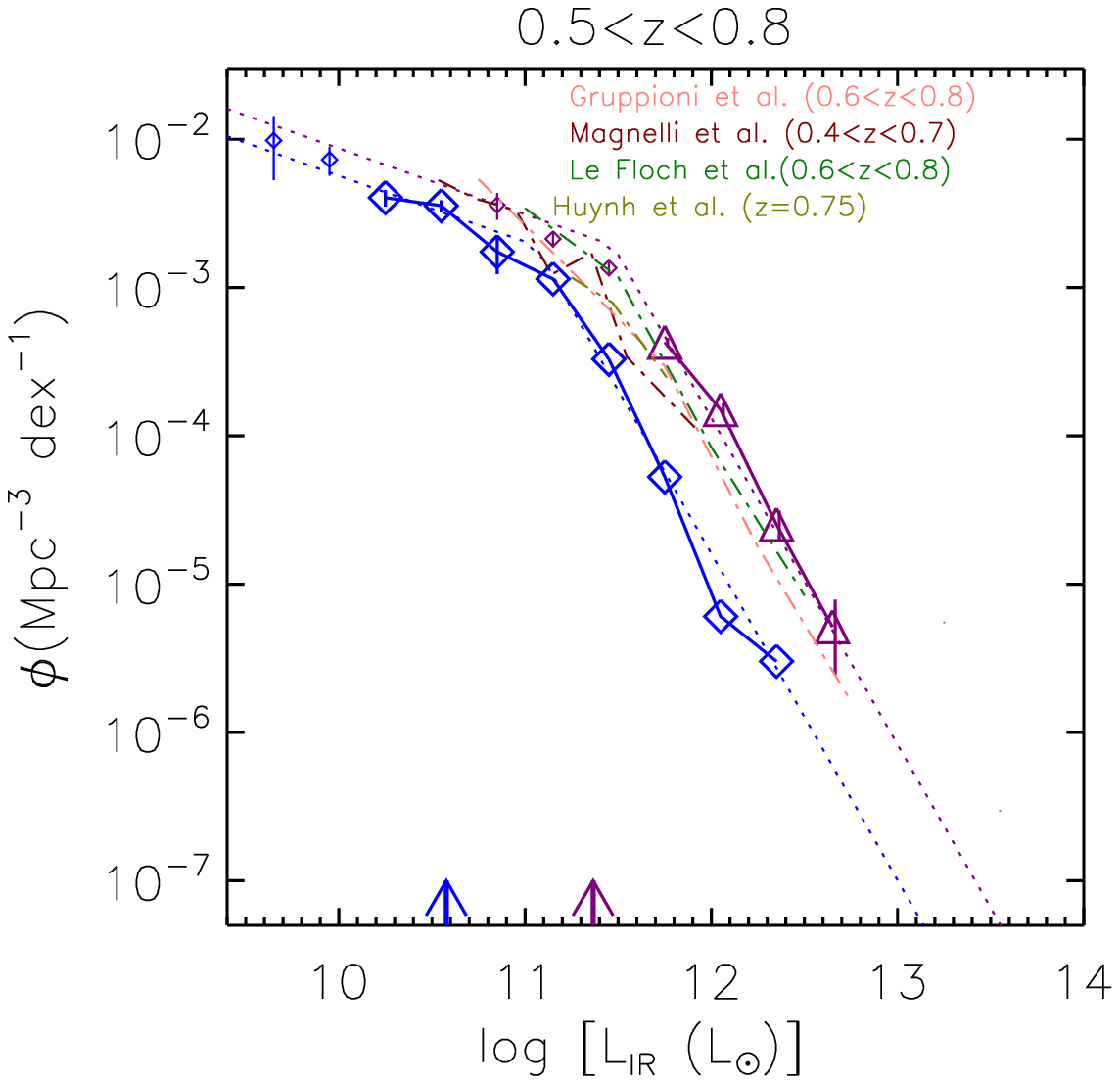}
 \includegraphics[height=7cm,trim={2cm 0 4cm 0},clip]{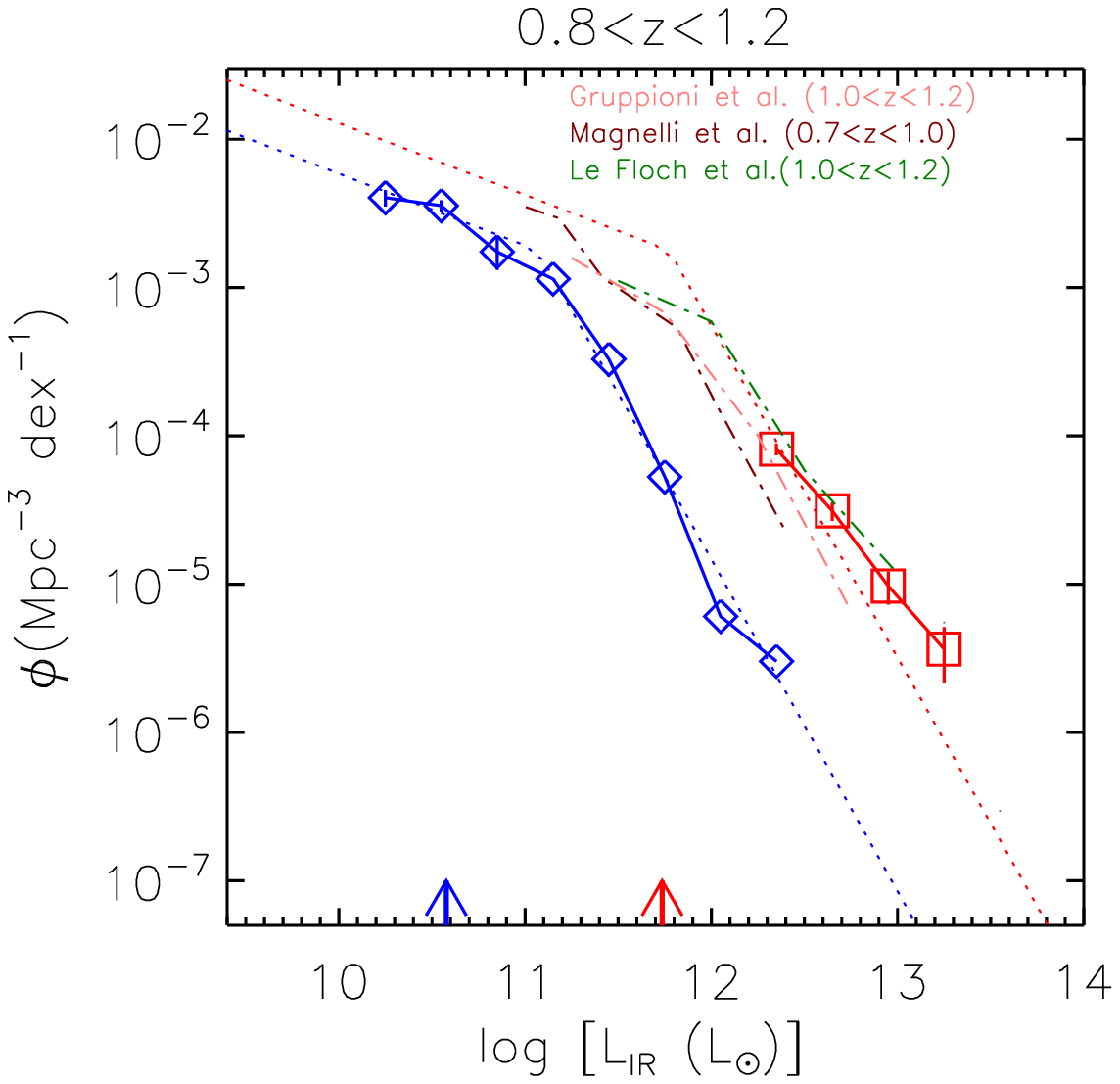}
 \includegraphics[height=7cm,trim={2cm 0 4cm 0},clip]{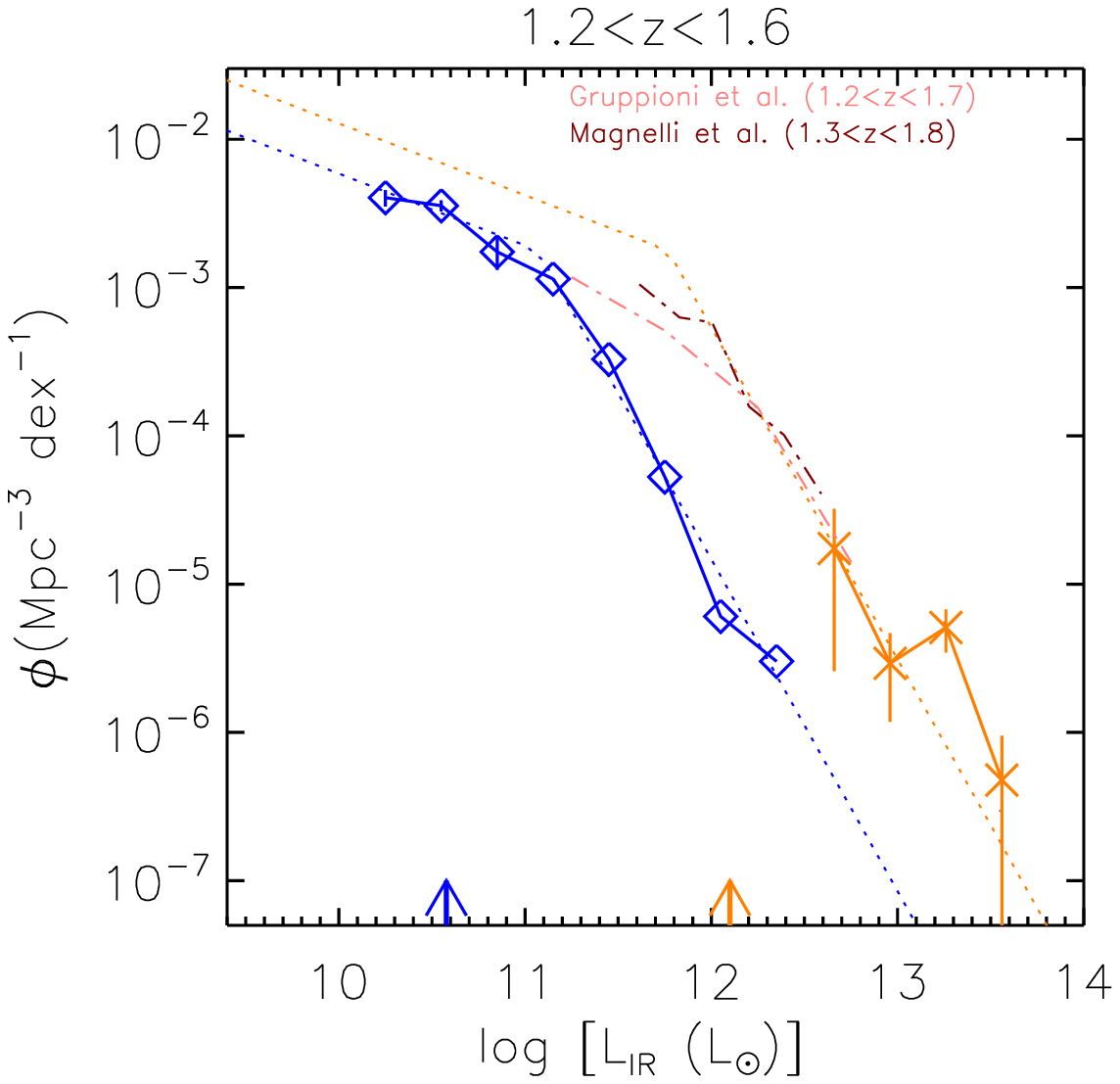}
 \end{center}
 \caption{
The TIR LFs from the SED fit.
The top panel shows results from all the redshift bins.
The succeeding panels show one redshift bin at a time for clarity.
 Vertical arrows show the luminosity corresponding to the flux limit at the central redshift in each redshift bin.
 The smaller data points at the faint ends are adopted from the NEP deep field, where AKARI data are deeper \citep{GotoCFHT}, and are included in the fit.
 We overplot z=0 IR LF based on the AKARI FIR all sky survey in the black dash-dot line \citep{Goto2011SDSS,2018MNRAS.474.5363K}.
 Overplotted previous studies are taken from 
\citet{2005ApJ...632..169L} in the dark-green, dash-dotted line,
 \citet{2013A&A...553A.132M}  in the dark-red,  dash-dotted line,  
 \citet{2007ApJ...667L...9H} in the dark-yellow,  dash-dotted line,
 \citet{2013MNRAS.432...23G}  in the pink,  dash-dotted line
 at several redshifts as marked in the figure.
 Best-fit parameters are presented in Table \ref{tab:fit_parameters}.
}
\label{fig:TLF}
\end{figure*}

 We take full advantage of 18-band  mid-IR coverage in SED-fitting to estimate $L_{\mathrm{TIR}}$.
   Although we have extensive photometric coverage in mid-IR, we caution readers that estimation of the $L_{\mathrm{TIR}}$ involves extrapolation to the far-IR wavelength range based on the SED models, and thus invites associated uncertainty, as we further discuss in Section 5.


Using photo-z,
we use the {\ttfamily  LePhare} code to find the best-fit SED to derive $L_{\mathrm{TIR}}$. 
Templates used are \citet{2003MNRAS.338..555L}.
Note that here, in addition to the AKARI's 9 mid-IR bands, we also used Spitzer \citep[IRAC1,2,3,4 and MIPS24,][]{2018ApJS..234...38N} and WISE ($W1,W2,W3,$ and $W4$) bands as well, i.e.,  we used 18 mid-IR bands in total.


 The $L18W$ flux \citep{2006PASJ...58..673M} are used to  apply the 1/$V_{\max}$  method, because it is a wide, sensitive filter (but using the $L15$ flux limit does not change our main results).
 We used \citet{2003MNRAS.338..555L}'s models for $k$-corrections to compute $V_{\max/mim}$.
We used redshift bins of 0.2$<z<$0.5, 0.5$<z<$0.8, 0.8$<z<$1.2,  and 1.2$<z<$1.6.  

 We show $L_{\mathrm{TIR}}$ LFs in Fig. \ref{fig:TLF}. For clarity, we separated LFs in four different panels at each redshift bin.
 For a local benchmark, we overplot one of the most accurate local IR LFs based on 15,638 IR galaxies from the AKARI all sky survey  \citep{2018MNRAS.474.5363K}.
 The TIR LFs show a strong evolution compared to local LFs, but again turns over at $z>1.2$.
 As in the case of 8$\mu$m LFs, the LF of the highest redshift bin is smaller, or comparable to the next bin at the lower redshift. 

It seems there are significant variations among previous studies plotted in various dash-dotted lines. AKARI's TIR LFs are at least consistent with one of the previous studies. Note that the redshift ranges are not completely matched, therefore, some variations are expected. Possible differences between our mid-IR based measurements and far-IR based measurements are further discussed in Section \ref{sec:discussion}.

\subsection{Total IR Luminosity density, $\Omega_{IR}$} \label{sec:madau}

Using LFs in previous sections, we next compute the IR luminosity density, to estimate the  cosmic star formation density \citep{1998ARA&A..36..189K}. 

 \subsubsection{Total IR Luminosity Density from  $L_{8\mu m}$ LFs} 
 First, we estimate Total IR Luminosity Density from  $L_{8\mu m}$ LFs.
 To do so, we need to convert $L_{8\mu m}$ to the total infrared luminosity.

$L_{8\mu m}$ and $L_{\mathrm{TIR}}$    have been reported  to correlate well \citep{2007ApJ...660...97C,2008A&A...479...83B}.
Using a large sample of 605 galaxies in the AKARI far-IR all sky survey,
\citet{Goto2011IRAS} derived the best-fit relation as
\begin{eqnarray}\label{eq:8um}
  L_{\mathrm{TIR}} (L_{\odot})= (20\pm5) \times \nu L_{\nu,8\mu m}^{0.94\pm0.01} (\pm 44\%).
\end{eqnarray}

$L_{\mathrm{TIR}}$ is from AKARI's far-IR photometry in 65, 90, 140, and 160 $\mu$m, and the $L_{8\mu m}$ measurement is from AKARI's 9$\mu$m flux.
Due to the improved statistics and the use of far-IR wavelengths (140 and 160$\mu$m),
this equation is superior to its precursors.

 The conversion, however, has been the largest source of error in estimating  $L_{\mathrm{TIR}}$ values from  $L_{8\mu m}$.  Reported dispersions are 37, 55 and 44\% by \citet{2008A&A...479...83B}, \citet{2007ApJ...660...97C}, and \citet{Goto2011IRAS}, respectively.  It should be kept in mind that the restframe $8\mu$m is sensitive to the star-formation activity, but at the same time, it is where the SED models have strong discrepancies due to the complicated  polycyclic aromatic hydrocarbon (PAH) emission lines. Possible SED evolution, and the presence of (unremoved) AGN will induce further uncertainty. A detailed comparison of different conversions is presented in Fig.12 of  \citet{2007ApJ...660...97C}, who reported a factor of $\sim$5 differences among various models.

 In addition, the above conversion is estimated using local star-forming galaxies, and thus, could be different for starburst or high redshift galaxies.

   For example, \citet{2012ApJ...745..182N} reported that the main-sequence galaxies 
 tend to have a similar  $L_{8\mu m}$/$L_{\mathrm{TIR}}$ regardless of $L_{\mathrm{IR}}$ and redshift, up to z$\sim$2.5, and  $L_{8\mu m}$/$L_{\mathrm{TIR}}$ decreases with increasing offset above the main sequence, possibly due to a change in the ratio of PAH to $L_{\mathrm{TIR}}$.
 \citet{2014A&A...566A.136M} also reported that  $L_{8\mu m}$/$L_{\mathrm{TIR}}$ is constant at below the main sequence, while it decreases with starburstiness at above the main sequence, concluding that starburst galaxies have deficient PAH emission compared with main-sequence galaxies. 
 Also \citet{2011A&A...533A.119E} showed that  $L_{8\mu m}$/$L_{\mathrm{TIR}}$ is different for starbursts.
  \citet{2018Kim} reported that  $L_{8\mu m}$/$L_{\mathrm{TIR}}$ with increasing $L_{\mathrm{TIR}}$
or increasing redshift up to z=0.9.
 A possible evolution with redshift was also discussed in \citet{2008ApJ...675..262R,2009ApJ...700..183H}.
   These results caution us that the single conversions from monochromatic IR luminosity to $L_{\mathrm{TIR}}$ are not likely to work at all redshifts for galaxies with different starburstiness.

In addition, \citet{2017ApJ...837..157S} reported metallicity and stellar mass dependence of  $L_{7.7\mu m}$/$L_{\mathrm{TIR}}$, indicating a paucity of PAH emission in low metallicity environments. They proposed separate  $L_{7.7\mu m}$/$L_{\mathrm{TIR}}$ conversions at above and below  log$(M_*/M_{\odot})=10$. 
 \citet{2014A&A...565A.128C} also found a weakening of the PAH emission in galaxies in low metallicities  and, thus, low stellar masses, suggesting PAH are destroyed in low metallicity environment by the UV radiation field which propagates more easily due to the lower dust content. The effect is most notable at  log$(M_*/M_{\odot})<8.5$, or at log$(M_*/M_{\odot})<9.5$.
 However, we note that most of our sample galaxies have log$(M_*/M_{\odot})>10$ and 12+log (O/H)$\sim$8.6 \citep{2017PASJ...69...70O}, where dependencies are less significant.

 On the other hand, \citet{2008A&A...479...83B}  stacked 24$\mu$m sources at $1.3<z<2.3$ in the GOODS fields to conclude that the correlation is valid to link $L_{8\mu m}$ and $L_{\mathrm{TIR}}$ at  $1.3<z<2.3$.
 \citet{Takagi_PAH} also showed that local $L_{7.7\mu m}$ vs $L_{\mathrm{TIR}}$ relation holds true for IR galaxies at z$\sim$1 (see their Fig.10).
 \citet{2008ApJ...675.1171P} showed that $z\sim$2 sub-millimeter galaxies lie on the relation between $L_{\mathrm{TIR}}$  and $L_{PAH,7.7}$ that has been established for local starburst galaxies. 
 The $S_{70}/S_{24}$ ratios of 70$\mu$m sources in \citet{2007ApJ...668...45P} are also consistent with the local SED templates.

 We further test this issue using our data in Section \ref{sec:discussion}.

 \subsubsection{Total IR Luminosity Density from  $L_{12\mu m}$ LFs} 
 $L_{12\mu m}$ is also reported to correlate with $L_{\mathrm{TIR}}$ \citep{2001ApJ...556..562C,2005ApJ...630...82P}.
 Due to the same reasons as  $L_{8\mu m}$ (improved statistics, and availability of 140 and 160$\mu$m), we use the following conversion \citep{Goto2011IRAS}. 


\begin{eqnarray}\label{eq:12um}
  L_{TIR}(L_{\odot}) = (17\pm4) \times \nu L_{\nu,12\mu m}^{0.96\pm0.01} (\pm 25\%).
\end{eqnarray}

This conversion agrees well with  the one given by \citet{1995ApJ...453..616S}.
We caution readers again here for the use of a single conversion for varieties of galaxies with different SFR at different redshifts. Results should be interpreted with this uncertainty in mind.

 \subsubsection{Integration to TIR density}

The derived total LFs are multiplied by $L_{\mathrm{TIR}}$ and integrated to measure the TIR density ( $\Omega_{\mathrm{TIR}}$).
We first fitted an analytic function to  integrate.



Following our previous work, we use a double-power law. With the lowest redshift LF, we first fit the normalization ($\Phi^{*}$) and slopes ($\alpha,\beta$). 
 At higher redshifts, statistics are  not enough  to  fit 4 parameters ($\Phi^{*}$, $L^*$, $\alpha$, and $\beta$) at the same time.  Therefore, we had to fix slopes and normalizations to those of the lowest redshift bin. Only $L^*$ is the free parameters at the higher-redshifts.
This is a common exercise with the limited depths of the current IR data \citep{2006MNRAS.370.1159B,2007ApJ...660...97C}.
Previous work also found a stronger evolution in luminosity than in density \citep{2005ApJ...630...82P,2005ApJ...632..169L}.

\begin{figure}
   \includegraphics[height=7.5cm,trim={2cm 0 1cm 0},clip]{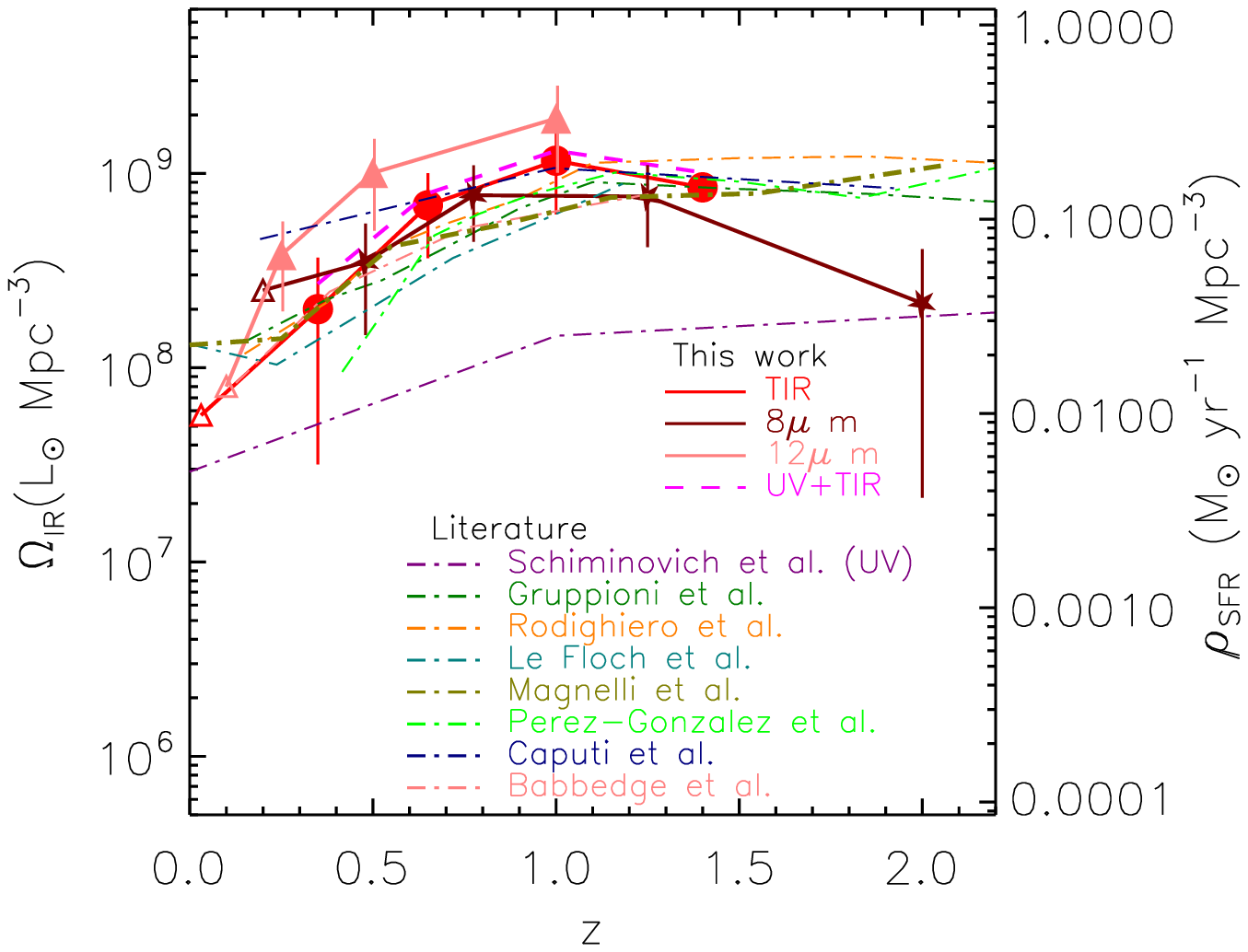}
 \caption{
 The evolution of the TIR luminosity density based on TIR LFs (red circles), 8$\mu$m LFs (stars), and 12$\mu$m LFs (filled triangles). 
Overplotted dot-dashed lines are estimates from the literature: 
\citet{2005ApJ...632..169L}, 
\citet{2009A&A...496...57M}, 
\citet{2005ApJ...630...82P}, 
\citet{2007ApJ...660...97C},   
\citet{2013MNRAS.432...23G},
\citet{2010A&A...515A...8R},
and \citet{2006MNRAS.370.1159B} are in cyan, yellow, green, navy, dark green, orange, and pink, respectively.
The purple dash-dotted line shows the UV estimate by \citet{2005ApJ...619L..47S}.
   The pink dashed line shows the total estimate of IR (TIR LF) and UV \citep{2005ApJ...619L..47S}. The open triangles are low-z results from \citet{2018MNRAS.474.5363K,2014ApJ...788...45T} in TIR,  8$\mu$m  and 12$\mu$m, respectively.
   Numerical values are presented in Table \ref{tab:madau_value}.  }
\label{fig:madau}
  \end{figure}

 Dotted-lines in Figs. \ref{fig:8um} to \ref{fig:TLF} show results of the fits.
 We integrate the double power laws outside the luminosity range to estimate  $\Omega_{\mathrm{TIR}}$.
Fig. \ref{fig:madau} shows $\Omega_{\mathrm{IR}}$ derived from the TIR LFs (red circles), 8$\mu$m LFs (brown stars), and 12$\mu$m LFs (pink filled triangles).

At the first glance,  $\Omega_{\mathrm{IR}}$ from 8$\mu$m and TIR LFs are consistent with each other. On the other hand,  $\Omega_{\mathrm{IR}}$ from 12$\mu$m LFs are $\sim$50\% larger, although error bars are touching each other. This could be due to the evolution on the $L_{12\mu m}/L_{8\mu m}$ ratio from local values. We further discuss this point in Section \ref{sec:discussion}. We also note that  $\Omega_{\mathrm{IR}}$ from 12$\mu$m is sensitive to the faint-end slope of 12$\mu$m LFs. In Fig. \ref{fig:12um}, we obtained steeper faint-end slopes than those of $L_{8\mu m}$ or $L_{IR}$ LFs. This is one of the reasons why  $\Omega_{\mathrm{IR}}$ from 12$\mu$m LFs are larger. However, even with AKARI's sensitivity, the observation might not be deep enough to reliably measure the faint-end slope of 12$\mu$m LFs, possibly because 12$\mu$m does not contain as luminous emission lines as in the case of 8$\mu$m.   Much deeper observations are awaited to clarify the issue.

Next, as indicated in LFs in previous sections,  $\Omega_{\mathrm{IR}}$ increases at $z<1.2$, then decreases at $z>1.2$.
Both 8$\mu$m and TIR LFs have shown the turnover at $z>1$.  
 Although this needs to be confirmed with more accurate data, we might have witnessed the turnover of the CSFH around z$\sim$2. This may be qualitatively consistent with previous reports by Herschel that the dust attenuation peaks and declines at $z>1.2$ \citep{2013MNRAS.432...23G,2013A&A...554A..70B}.

 This is an interesting implication, but it is unfortunate that our error bars are too large  to draw significant conclusions. As we mentioned in the introduction, mid-IR is a relatively unexplored wavelength range. At the same time, however, it means that mid-IR has a great room to be explored. Only 65cm diameter telescope, AKARI, revealed comparable results on $\Omega_{\mathrm{IR}}$ to those from 3m Herschel telescope in far-IR. This is because mid-IR detectors are still more sensitive than far-IR. If larger aperture mid-IR telescopes become available in the future, such as SPICA \citep{2018PASA...35...30R} and James Webb Space Telescope \citep{2006SSRv..123..485}, mid-IR is a good wavelength range to invest, having extinction-free advantage in IR, and yet more sensitive than far-IR (for typical SEDs).

\section{Discussion}\label{sec:discussion}

 In the previous section, in addition to $L_{TIR}$ measurement, we converted $L_{8\mu m}$ and $L_{12\mu m}$ into  $L_{TIR}$. 
 However, the conversions are based on local star-forming galaxies. It is uncertain whether it holds at higher redshift or not,   including starburst galaxies.
 The conversion using a single relation might be too simple, in the presence of multiple components of dust at different temperatures, with different star-formation rates, and metallicities.

\begin{figure}
     \includegraphics[height=7.5cm,trim={2cm 0 1cm 0},clip]{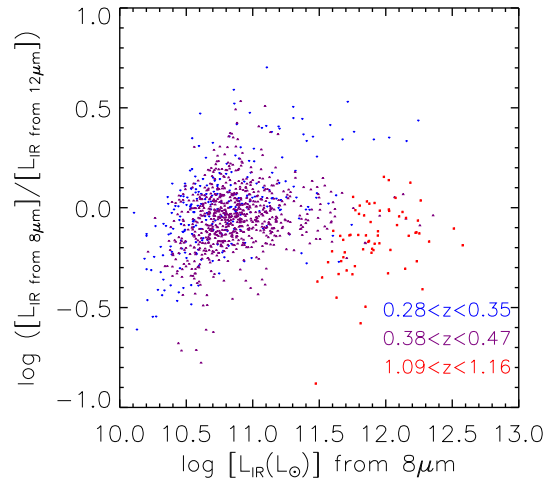}
\caption{Comparison of $L_{TIR}$ estimated from $L_{8\mu m}$ and $L_{12\mu m}$ using equations \ref{eq:8um} and \ref{eq:12um}, respectively, in three overlapping redshift ranges between 8$\mu$m and 12$\mu$m LFs redshift bins. Median offset for the highest redshift bin is dex -0.14. 
\label{fig:8vs12}}
  \end{figure}

 Following the results in the literature discussed in Section 4.3, in this section, we compare  $L_{TIR}$ estimated from $L_{8\mu m}$ and $L_{12\mu m}$ from equations \ref{eq:8um} and \ref{eq:12um} in three overlapping redshift ranges in Fig.\ref{fig:8vs12} using our data. One can immediately notice that the relation deviates at log$L_{TIR}>$12 (or equivalently at $z>1$). The median offset is the largest for the highest redshift bin with dex $-$0.14. Therefore,   we caution readers that the conversions in  equations \ref{eq:8um} and \ref{eq:12um} may not be valid at  log$L_{TIR}>$12, and that the possible inconsistency in Figs. \ref{fig:TLF}  and \ref{fig:madau} between mid- and far-IR measurements could be the result of the change in the SED, rather than incorrect measurements on either. In this sense, the mid- and far-IR measurements should be complimentary, and are both important. Having AKARI's superior mid-IR data, it is our important task to exploit both data to investigate mid-to-far IR SED evolution, and reveal physical origins behind them. 
 Our attempt on this is in preparation (Kim et al. in preparation).

 \section{Summary} \label{sec:sum}

 Previously AKARI NEP wide field lacked deep optical photometry, and thereby, accurate photo-z, despite the presence of space-based 9-band mid-IR photometry from AKARI.
 To rectify the situation,   we have obtained deep optical 5-band imaging covering the entire 5.4 deg$^2$ of the NEP wide field, using the new Hyper Suprime-Cam mounted on the Subaru 8m telescope. Combined with the CFHT $u$-band imaging we have also taken,
 for the first time, we used  all of the AKARI's data over the 5.4 deg$^2$.

 In addition to AKARI's continuous 9-band mid-IR filter coverage (2.4, 3.2, 4.1, 7, 9, 11, 15, 18, and 24$\mu$m), we combined with archival WISE (4 bands) and Spitzer (5 bands) data.
 In total, we have the 18-band mid-IR photometry, which is the most complete mid-IR photometry to date for thousands of galaxies.

    We presented restframe 8$\mu$m, 12$\mu$m LFs at 0.35$<$z$<$2.2.
  We also estimated total infrared LFs through SED fitting to the 18-band mid-IR data.  
  Thanks to the large area coverage, the bright-ends are better-determined.
The resulting LFs are consistent with our previous work \citep{GotoTakagi2010,GotoCFHT}, but with much reduced statistical errors thanks to  the new HSC and AKARI data. It is interesting to note that $\Omega_{IR}$ becomes smaller at $z>1.2$, possibly suggesting the turnover of the CSFH.

Until recently the mid-IR SED studies were limited to a small number of bright galaxies with mid-IR spectra (e.g., Spitzer IRS). Our work demonstrated that such studies can be done with photometry only, once enough filter coverage such as AKARI's becomes available, paving the way to statistical studies of mid-IR SEDs in the future.



%

%
%
%
%
%
%
%
%
%



\begin{table*}
 \centering
 \begin{minipage}{180mm}
  \caption{Best fit parameters for 8,12$\mu$m and TIR LFs}\label{tab:fit_parameters}
  \begin{tabular}{@{}ccccllcccc@{}}
  \hline
   Redshift & LF & $L^*$ ($ 10^{10} L_{\odot}$)& $\Phi^*(10^{-3} \mathrm{Mpc^{-3} dex^{-1}})$ & $\alpha$ & $\beta$  \\ 
 \hline
 \hline
0.28$<$z$<$0.47  &   8$\mu$m & $1.8^{+0.1}_{-0.2}$ & 1.6$^{+0.2}_{-0.1}$ & 1.56$^{+0.03}_{-0.24}$   &   2.6$^{+0.1}_{-0.1}$ 	\\
0.65$<$z$<$0.90  &   8$\mu$m & $3.3^{+0.2}_{-0.2} $ & $1.6 $  &1.56 & 2.6	\\
1.09$<$z$<$1.41    &   8$\mu$m & $4.1^{+0.2}_{-0.2} $ & $1.6 $  &1.56 & 2.6	\\  
1.78$<$z$<$2.22    &   8$\mu$m & $1.4^{+0.3}_{-0.2} $ & $1.6 $  &1.56 & 2.6	\\  
 \hline
0.15$<$z$<$0.35  &   12$\mu$m & $8.2^{+4.0}_{-0.6} $ &  1.5$^{+0.3}_{-0.3}$ &1.9$^{+0.1}_{-0.1}$   &   2.8$^{+0.5}_{-0.1}$ \\
0.38$<$z$<$0.62  &   12$\mu$m & 18$^{+1}_{-2} $ &  1.5 & 1.9   &  2.8 \\
0.84$<$z$<$1.16  &   12$\mu$m & 32$^{+2}_{-1} $ &  1.5 & 1.9   &  2.8 \\
   %
 \hline
0.2$<$z$<$0.5  &  Total  & 5.3$^{+1.1}_{-1.1} $& 2.9$^{+0.3}_{-0.3}$ &  1.4$^{+0.1}_{-0.2}$   &   2.7$^{+0.4}_{-0.4}$ 	\\
0.5$<$z$<$0.8  &  Total  & 17$^{+1}_{-2} $ &2.9 &  1.4   &   2.7	\\
0.8$<$z$<$1.2  &  Total  & 28$^{+2}_{-2} $ &2.9 &  1.4   &   2.7	\\
1.2$<$z$<$1.6  &  Total  & 20$^{+6}_{-9} $ &2.9 &  1.4   &   2.7	\\
 \hline
\end{tabular}
\end{minipage}
\end{table*}

\begin{table}
 \centering
 \begin{minipage}{180mm}
  \caption{TIR luminosity density as a function of redshift as in Fig. \ref{fig:madau}.}\label{tab:madau_value}
  \begin{tabular}{@{}ccclllcccc@{}}
  \hline
 z &    $\Omega_{\mathrm{TIR}}(L_\odot Mpc^{-3}/10^8)$\\ 
 \hline
0.35 &  2.1$\pm$ 1.8 \\
0.65 &  6.8$\pm$ 3.2 \\
1.00 & 11.2$\pm$ 5.3 \\
1.40 &  7.9$\pm$ 0.9 \\
   \hline
    z &    $\Omega_{TIR_{8\mu m}}(L_\odot Mpc^{-3}/10^8)$ \\ 
 \hline
0.48 &  3.5$\pm$ 1.9\\
0.77 &  7.6$\pm$ 3.2\\
1.25 &  7.5$\pm$ 3.4\\
2.00 &  2.1$\pm$ 1.9\\
\hline
        z &    $\Omega_{TIR_{12\mu m}}(L_\odot Mpc^{-3}/10^8)$ \\
 \hline
0.25 &  2.6$\pm$ 1.4\\
0.50 &  6.6$\pm$ 1.3\\
1.00 & 12.1$\pm$ 2.4\\       
       \hline
  \end{tabular}
\end{minipage}
\end{table}



\begin{ack}
 We thank the anonymous referee for many insightful comments, which significantly improved the paper.
 We are grateful for Tina Wang, and Simon Ho for careful proof-reading of the paper.
 TG acknowledges the support by the Ministry of Science and Technology of Taiwan through grant 105-2112-M-007-003-MY3.
 MI acknowledges the support from the grant No. 2017R1A3A3001362 of the National Research Foundation of Korea (NRF).
 TM is supported by NAM-DGAPA PAPIIT IN104216, IN111379 and CONACyT Grant 252531.
 YO  acknowledges the support by the Ministry of Science and Technology of Taiwan through grant MOST 107-2119-M-001-026.
\end{ack}

%
%


\bibliography{201404_gotob} 
\bibliographystyle{myaasjournal}

\end{document}